\documentclass[structabstract]{aa}  
\usepackage{natbib}
\usepackage{graphicx}
\usepackage{url}
\usepackage{txfonts}
\usepackage[T1]{fontenc}
\usepackage[latin9]{inputenc}
\begin{document}

\def\teff{$T\rm_{eff }$}
\def\kms{$\mathrm {km s}^{-1}$}
\def\gtsima
{\hbox{\raise0.5ex\hbox{$>\lower1.06ex\hbox{$\kern-1.07em{\sim}$}$}}}
\def\ltsima
{\hbox{\raise0.5ex\hbox{$<\lower1.06ex\hbox{$\kern-1.07em{\sim}$}$}}}

\title{High Resolution HDS/SUBARU chemical abundances \\
of the young stellar cluster Palomar 1}

   \author{L. \,Monaco\inst{1}, 
I. \,Saviane \inst{1},
M. \,Correnti \inst{2},
P. \,Bonifacio\inst{3,4}, 
\and 
D. \,Geisler \inst{5}
          }

\institute{
European Southern Observatory, Casilla 19001, Santiago, Chile
\and
Istituto Nazionale di Astrofisica --
Osservatorio Astronomico di Bologna, 
Via Ranzani 1, 40127, Bologna, Italy
\and
GEPI, Observatoire de Paris, CNRS, Universit\'e Paris Diderot ; Place Jules
Janssen, 92190 Meudon, France
\and
Istituto Nazionale di Astrofisica --
Osservatorio Astronomico di Trieste, Italy
via G. B. Tiepolo 11 34143 Trieste, Italy
\and
Universidad de Concepci\'on,
Casilla 160-C, Concepci\'on, Chile
}

\authorrunning{Monaco et al.}
\mail{lmonaco@eso.org}

\titlerunning{Pal1 chemical abundances}

\date{Received ...; Accepted...}

\abstract
{Palomar\,1 is a peculiar globular cluster (GC). It is the youngest Galactic GC
and it has been tentatively associated to several of the substructures recently
discovered in the Milky Way (MW), including the Canis Major (CMa) overdensity
and the Galactic Anticenter Stellar Structure (GASS).}
{In order to provide further insights into its origin, we present the
first high resolution chemical abundance analysis for one red giant in Pal\,1.} 
{We obtained high resolution (R=30000) spectra for one red giant star in Pal\,1
using the High Dispersion Spectrograph (HDS) mounted at the SUBARU telescope. 
We used ATLAS-9 model atmospheres coupled with the SYNTHE and WIDTH calculation
codes to derive chemical abundances from the measured line equivalent widths of
18 among $\alpha$, Iron-peak, light and heavy elements.} 
{The Palomar~1 chemical pattern is broadly compatible to that of the MW open
clusters population and similar to disk stars. It is, instead, remarkably
different from that of the Sagittarius (Sgr) dwarf spheroidal galaxy.} 
{If Pal\,1 association with either CMa or GASS will be confirmed, this will
imply that these systems had a chemical evolution similar to that of the
Galactic disk.}

\keywords{Stars: abundances, Galaxy: abundances, (Galaxy:) globular clusters:
individual: Palomar\,1}

\maketitle

\section{Introduction \label{sec:Introduction}}

Palomar 1 is sparsely populated and the youngest among Galactic globular
clusters \citep[$M_{V}=-2.5\pm0.5$; age=$6.3$---$8$~Gyr,][hereafter
R98a]{rosenberg_etal98a}. Very little is known on this cluster:
\citet[][hereafter R98b]{rosenberg_etal98b} provided the only spectroscopic
study of Pal\,1 red giant branch stars to date. By observing four stars in the
Calcium II Triplet infrared region, they derived a systemic heliocentric
velocity of $v_{\rm helio}$=-82.8$\pm$3.3\,\kms and a metallicity
[Fe/H]=-0.71$\pm$0.20. Besides its unusual age and metallicity, the global slope
of the Pal1\, mass function (MF) is flatter than Galactic globular clusters
(GCs) at similar locations and, hence, it does not fit the correlation with
Galactic position ($R_{GC}$, $Z_{GP}$), and metallicity {[}Fe/H{]} followed by
old Milky Way (MW) GCs \cite[see Fig.\,12 in R98a and][]{djorgovski93}. R98a
considered also the possibility of Pal\,1 being an open cluster (OC). The
concentration derived, c=1.6, higher than any OC, and its position in the Galaxy
(R$_{GC}$=17.3, Z=3.6\,kpc) would make it a peculiar OC, though.  Its mass and
half light radius place Pal\,1 among the ``lucky survival" in the vital diagram
of GCs \citep[][]{gnedin97}. This suggests that Pal\,1 has lost an important
fraction of its mass. In fact, R98a calculate an evaporation time of 0.74\,Gyr
and concluded that Pal1\, might be on the verge of destruction.

Several authors suggested that young GCs like Pal\,l may be related to dwarf
satellites disrupted in the early  stages of formation of the MW \citep[see,
e.g.,][]{lynden-bell} and it is now clear that accreted satellites might have in
fact contributed to the star clusters population of the Galactic halo and likely
also to the disk OCs population \citep[e.g.][]{frinchaboy04,carraro09,law10}. 

Indeed, putative association of Pal\,1 with several of the substructures
discovered in the Galactic halo over the past decade has been suggested. In
particular, association of Pal\,1 with the ``GASS" \citep[Galactic Anticenter
Stellar Structure.][]{crane03,frinchaboy04} and the controversial Canis Major
dSph galaxy \citep[][]{momany06,martin04,forbes10} have been proposed.
Association with the ``orphan stream" \citep[][]{belokurov07} was also
suggested, although a recent study of the stream orbit argue against this
possibility \citep[][]{newberg10}.  While Pal\,1 seems not to be related with
the Sagittarius (Sgr) dwarf spheroidal galaxy tidal streams \citep[][]{law10}
either, it is interesting to notice that Pal\,1 comfortably fits into the Sgr
age {\it vs} metallicity relation (AMR) derived by \citet[][]{siegel07}. Pal\,1
and the young Sgr GCs Ter\,7 and Pal\,12, are also the only three GCs having
[Fe/H]$>$-1.2 and lying at Galactocentric distances R$_{GC}>$8\,kpc. 

Recently, \citet[][]{no10} detected tidal tails out to a distance of about
90\,r$_h$ ($\sim$1 degree) from either side of Pal\,1. They argue that Pal\,1
might have been evolving in a dwarf galaxy accreted by the MW less than 500\,Myr
ago.

Chemical abundance patterns provide information about the formation history of
the system in which stars were formed. In particular, the MW satellites and
their GCs systems present chemical compositions remarkably different from what
observed in the MW \citep[see,
e.g.][]{venn04,monaco05,lanfranchi06,monaco07,sbordone07}. In order to shed
light onto its origin, we present here the first high resolution chemical
abundance analysis for a red giant star in the peculiar GC Palomar\,1.

\section{Observations and data reductions}

We observed the Red Giant Branch (RGB) star Pal1-I (according to the
nomenclature introduced in R98b) using the High Dispersion Spectrograph
\citep[HDS,][]{noguchi02} mounted at the SUBARU telescope in Mauna Kea (Hawaii).
The Pal\,1 RGB contains very few stars (see, e.g., Fig.\,1 in R98b) and,
although more stars were in our target list, only this one could be eventually
observed. While having only one star may be a concern, the homogeneity of the
iron abundances of all stars measured by R98b suggest that Pal\,1-I is likely
not a peculiar object. Yet, the presence of a spread in the measured abundance
ratios is an issue which cannot be addressed in the present analysis.

Two 1~hr exposures were taken on November 2nd 2007 in service mode
(Progr.ID:~S07A-144S) using the standard StdYb setup which covers the spectral
range  $\sim$4100-6900\,\AA. For each exposure, the entire spectral range is
recorded in two frames through a blue and a red CCD. We adopted a 1.2\arcsec
slit width which provides a spectral resolution of R=30000.

We employed the overscan region to perform the bias subtraction with the aid of
in-house script available from the HDS/SUBARU web
page\footnote{\url{http://www.naoj.org/Observing/Instruments/HDS/hdsql-e.html}}.
To remove cosmic rays we used IRAF\footnote{IRAF is distributed by the National
Optical Astronomy Observatories, which is operated by the association of
Universities for Research in Astronomy, Inc., under contract with the National
Science Foundation.} scripts available at the same URL. We applied a median
filter to the data and compared the counts to the original frames. When high
peaks were detected the pixels values were substituted with the counts of the
median-filtered image. The procedure is described in \citet[][]{aoki05}.

Data reduction  was performed using standard IRAF tasks and included order
tracing, flat-fielding, background and sky subtraction and extraction. Standard
Th-Ar arcs were used for wavelength calibrations. The four scientific frames
were reduced individually. 

\begin{figure}
\includegraphics[width=1\columnwidth]{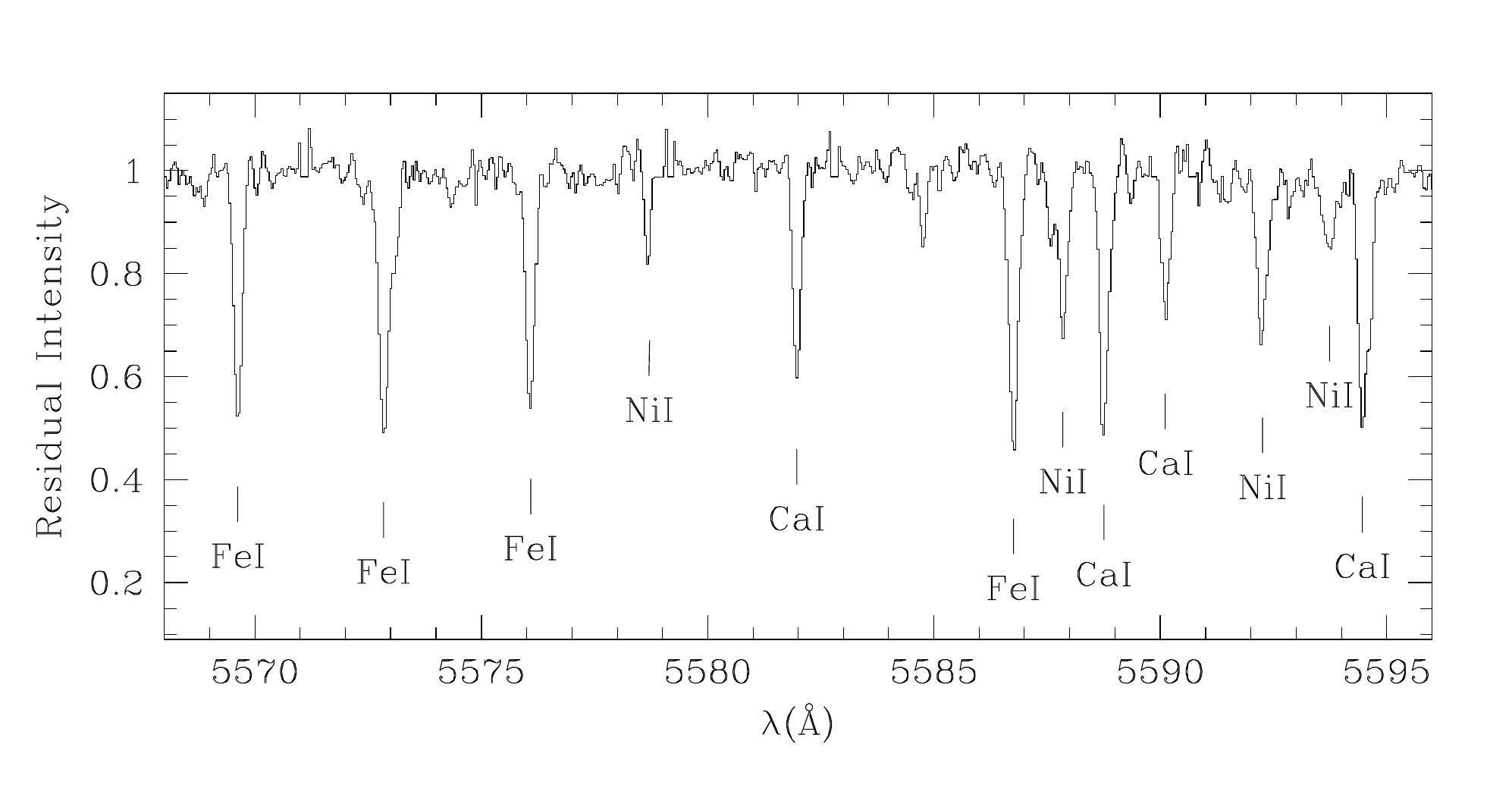}
\caption{Sample of the spectrum obtained for Pal1-I. Lines corresponding to
different elements are marked in the figure.}\label{spec}
\end{figure}

The object radial velocity (RV) was computed for each frame by cross-correlating
the observed spectra with a synthetic one of similar atmospheric parameter using
the IRAF task {\tt fxcor}. Finally, the single exposure spectra were reduced at
rest frame and averaged. The final spectrum has a signal to noise of$\sim$40 at
560~nm. The IRAF task {\tt rvcorrect} was used to calculate the earth motion and
convert the observed RV to the heliocentric system. Fig.\,\ref{spec} shows a
sample of the obtained spectrum.

\begin{table*}
\caption{Basic parameters of Pal1-I. Measured radial velocity and spectra signal-to-noise
ratios are also reported.}

\label{star} 

\centering{}\begin{tabular}{ccccccccc}
\hline 
$\alpha$(J2000)  & $\delta$(J2000)  & V  & V-I  & E(B-V)  & S/N@560nm  & v$_{helio}$ (\kms)  & v$_{lsr}^{a}$ (\kms)  & v$_{gsr}^{a}$ (\kms) \tabularnewline
\hline 
03:33:21.896  & +79:35:38.76  & 16.39  & 1.16  & 0.15  & 40  & -76.1$\pm$0.5  & -71.7  & +87.5\tabularnewline
\hline
\end{tabular}\\
 \smallskip{}
 $^{a}$ Velocity in the local and galactic standard of rest are computed
according to the following relations \citep[see][]{palma00}:\\
 $v_{lsr}=v_{helio}+(9\, cos\, b\, cos\, l\,+11\, cos\, b\, sin\, l\,+6\, sin\, b)$;
 $v_{gsr}=v_{lsr}+(220\, cos\, b\, sin\, l)$ %$$

\end{table*}

Table~\ref{star} reports Pal1-I coordinates, magnitudes from R98b, S/N and the
measured radial velocity. The reported RV (-76.1\,\kms) is the average of the
measures in the 4 frames available. The standard deviation of the four derived
RVs is $\sim$0.2~\kms. HDS is located at a Nasmyth platform and, as such, should
not be critically affected by flexures. However, by measuring a few sky lines we
evaluate a residual  uncertainty in the RV of the order of an additional
$<$0.2~\kms. We adopted a conservative 0.5~\kms estimate for the RV error. For
Pal1-I, R98b obtained  v$_{helio}$=-81.0$\pm$4.6\,\kms from lower resolution
spectra (R$\simeq$4000) centered on the Ca~II triplet. Clearly, the two measures
are compatible with each other within 1---$\sigma$.

We remark here that R98a also acquired control fields and showed that the region
of color-magnitude diagram (CMD) in which Pal\,1-I is selected is virtually free
from field stars contamination (see Fig.\,5 in R98a). This is confirmed by the
Galaxy model of 
\citet[][]{robin03}\footnote{\url{http://model.obs-besancon.fr/}} which
predicts  1$\times$10$^{-3}$ stars in a 80\arcsec\, region centered on Pal\,1
(Pal\,1-I lies at about 49\arcsec from the cluster center) in a CMD box defined
by  1.11$<$(V-I)$<$1.21; 16.09$<$V$<$16.69 and having radial velocity in the
range -80$<$v$_{helio}$(\kms)$<$-70. The above figure raises to
7$\times$10$^{-2}$ if we consider the whole cluster region out to the tidal
radius (r$_t$=525\arcsec, R98a). Therefore, Pal\,1-I is unlikely to be a field
star.

\section{Chemical abundance analysis}

\subsection{Linelist and equivalent width measurements}

Species abundances are obtained from line equivalent widths (EWs) measured using
the DAOSPEC code. The Gaussian approximation adopted by DAOSPEC represents a
reliable approximation to the line profile at the resolution under consideration
up to relatively strong lines \citep[$\sim$150~m\AA\,,
see][]{daospec,pancino10}. 

Appendix~\ref{appendix} reports the adopted linelist, atomic parameters and
measured line EWs. Atomic parameters  were retrieved from the Vienna Atomic Line
Database
\citep[VALD\footnote{\url{http://vald.astro.univie.ac.at/~vald/php/vald.php?docpage=usage.html}},][]{vald}
with the exception of the single lines measured for Mn and Co. Both these
elements are known to be significantly affected by hyperfine splitting. For
these transitions we adopted the hyperfine structures (HFS) tabulated by
\citet[][]{prochaska00}.

In order to perform a differential solar analysis, we broadened the
\citet[][]{solarflux} Solar Flux
Atlas\footnote{\url{http://kurucz.harvard.edu/sun.html}} to the resolution of
our SUBARU Pal1-I spectrum (R=30000) and measured the solar line EWs (see
appendix~\ref{appendix}) as above.

Silicon solar abundances calculated adopting VALD log~gf present a large
dispersion. This may imply an important level of internal inconsistency among
these log~gf. Therefore, for Si lines we adopted the oscillator strengths of
\citet[][]{bensby09}.

\subsection{Model Atmospheres and chemical abundances}

An initial estimate of the atmospheric parameters was obtained from the R98b
Pal1-I photometry (see Table~\ref{star}). Adopting a reddening of E(B-V)=0.15
\citep[][]{harris}, we obtain an effective temperature of \teff=4820~K from the
\citet[][]{alonso} calibrations for the Pal1-I (V-I) color. We note that the
\citet[][hereafter SFD98]{cobe} reddening maps provide a significantly larger
reddening value,  E(B-V)=0.19, which implies \teff=4950~K. However, by applying
the \citet[][]{bonifacio00} correction to SFD98 reddening values larger than
E(B-V)=0.10, we obtain E(B-V)=0.16 and, correspondingly, \teff=4850~K.

Adopting ages between 4 and 8 Gyr and a metallicity of Z=0.004
\citep[][]{rosenberg_etal98a}, we derived a gravity of $\log g = 2.4\pm 0.5$~dex
by comparison with the \citet[][]{leo} and the \citet[][]{basti} isochrones.

A first model atmosphere having the above parameters and [M/H]=-0.5 was computed
using the Linux port of version 9 of the ATLAS code and abundances were derived
from the line EWs using the WIDTH code \citep[][]{k93,sbordone04}. Iron lines
were used to refine the atmospheric parameters. The microturbulence velocity
($\xi$) was estimated by minimizing the dependence of the abundances from the
measured EWs. The effective temperature is determined by imposing that
abundances should be independent from the transition excitation potential.  We
finally adopt a slightly hotter temperature of \teff=5000~K for Pal1-I. Iron
abundances derived by Fe~I and Fe~II lines provide an excellent agreement,
confirming the photometric gravity as a proper value.

For the Sun, Fe abundances calculated adopting a model atmosphere having [M/H]=0
and the standard solar effective temperature and gravity (\teff=5777~K, $\log g
= 4.44$ ) satisfy well the above requirements, while for the microturbulence we
obtain $\xi$=0.9\,\kms, well within the range of figures adopted in the
literature \citep[see for instance][]{prochaska00,bensby09}.

\begin{table}
\caption{Atmospheric parameters adopted for Pal1-I and the Sun.}

\label{PA} 

\centering{}\begin{tabular}{lcccc}
\hline 
Star  & \teff  & log g  & $\xi$  & {[}M/H{]}\tabularnewline
 & K  &  & \kms  & \tabularnewline
\hline 
\object{Pal1-I}  & 5000  & 2.40  & 1.0  & $-0.5$ \tabularnewline
\object{Sun}  & 5777  & 4.44  & 0.9  & $0.0$ \tabularnewline
\hline
\end{tabular}
\end{table}

Table~\ref{PA} summarizes the atmospheric parameters adopted for Pal1-I and the
Sun. Model atmospheres having these parameters were calculated and employed
within the WIDTH code to derive the species abundances from the measured EWs for
all elements but Mn and Co. For these elements the abundances were obtained by
comparing the measured EWs to that of synthetic lines calculated with the SYNTHE
code along with the HFS presented in \citet[][]{prochaska00}. The abundances
obtained for each line are reported in appendix~\ref{appendix}. Table~\ref{abun}
presents the average abundances obtained for each species both for the Sun and
Pal1-I. In the last two column we report the solar abundances of
\citet[][hereafter GS98]{gs98} and the difference with the present analysis.
While for several elements we obtain a good agreements with GS98, we also find
notable discrepancies in a few elements. This is not uncommon in differential
abundance analysis \citep[see,
e.g.,][]{prochaska00,bensby03,bensby09,pancino10}.  In particular, we find high
discrepancies in the derived Al (-0.25~dex) and Y (+0.43~dex) abundances.  The
two Al lines measured here (6696~\AA and 6698~\AA) are known to give
systematically lower abundances and, in fact, several authors adopt
astrophysical values of the oscillator strengths \citep[see, e.g.,][hereafter
B03]{bensby03}. We measured only one Y line, which is also known to provide
abundances higher than other lines (see B03). The two La lines measured in the
Pal1-I spectrum were not detected in the Sun. As such, for La we adopt the GS98
solar value.

\begin{table*}
\caption{Average elemental abudances measured for both Pal1-I and the Sun.
The gaussian dispersion ($\sigma$) around the mean and the number
of lines used are also reported. The standard deviation of the mean
can be obtained as $\sigma$/$\sqrt{(}lines)$. The last two columns
report the GS98 solar values and the difference with respect to our
measurements.}

\label{abun} 

\centering{}\begin{tabular}{ll|cccr|crr|cc}
\hline 
Element  & Ion  & $\epsilon$  & {[}X/Fe~I{]}  & $\sigma$ & lines  & $\epsilon$  & $\sigma$ & lines  & $\epsilon(Sun)$  & {[}X/H{]}$_{Sun}$\tabularnewline
 &  & \multicolumn{4}{c|}{Pal1-I} & \multicolumn{3}{c|}{Sun} & GS98  & t.w.-GS98 \tabularnewline
\hline 
Al  & I  & 5.98  & +0.25  & 0.02 & 2  & 6.22  & 0.04  & 2  & 6.47  & -0.25 \tabularnewline
Ba  & II  & 2.00  & +0.24  & 0.02 & 2  & 2.25  & 0.07  & 2  & 2.13  & +0.12 \tabularnewline
Ca  & I  & 5.93  & +0.04  & 0.14 & 8  & 6.38  & 0.13  & 6  & 6.36  & +0.02 \tabularnewline
Co  & I  & 4.42  & -0.06  & ---- & 1  & 4.97  & ----  & 1  & 4.92  & +0.05 \tabularnewline
Cr  & I  & 4.93  & -0.23  & 0.20 & 5  & 5.65  & 0.07  & 4  & 5.67  & -0.02 \tabularnewline
Fe  & I  & 7.09  & {[}FeI/H{]}=-0.49  & 0.18 & 108 & 7.58  & 0.12  & 95  & 7.50  & +0.08 \tabularnewline
Fe  & II  & 7.08  & {[}FeII/H{]}=-0.53  & 0.17 & 7  & 7.61  & 0.06  & 7  & 7.50  & +0.11 \tabularnewline
La  & II  & 0.93  & +0.29  & 0.05 & 2  & 1.13  & ----  & 0  & 1.13  & ----- \tabularnewline
Mg  & I  & 7.11  & +0.11  & 0.08 & 3  & 7.49  & 0.17  & 3  & 7.58  & -0.09 \tabularnewline
Mn  & I  & 4.63  & -0.22  & ---- & 1  & 5.34  & ----  & 1  & 5.39  & -0.05 \tabularnewline
Na  & I  & 6.07  & +0.38  & 0.12 & 2  & 6.18  & 0.06  & 2  & 6.33  & -0.15 \tabularnewline
Ni  & I  & 5.76  & -0.03  & 0.17 & 25  & 6.28  & 0.16  & 24  & 6.25  & +0.03 \tabularnewline
Sc  & II  & 2.81  & -0.01  & 0.14 & 5  & 3.31  & 0.08  & 4  & 3.17  & +0.14 \tabularnewline
Si  & I  & 7.04  & -0.01  & 0.10 & 7  & 7.54  & 0.09  & 7  & 7.55  & -0.01 \tabularnewline
Ti  & I  & 4.57  & +0.10  & 0.18 & 17  & 4.96  & 0.09  & 12  & 5.02  & -0.06 \tabularnewline
V  & I  & 3.42  & -0.06  & 0.15 & 8  & 3.97  & 0.19  & 7  & 4.00  & -0.03 \tabularnewline
Y  & II  & 1.86  & -0.32  & ---- & 1  & 2.67  & ----  & 1  & 2.24  & +0.43 \tabularnewline
Zn  & I  & 4.40  & +0.38  & ---- & 1  & 4.51  & ----  & 1  & 4.60  & -0.09 \tabularnewline
\hline 
 &  &  &  &  &  &  &  &  &  & \tabularnewline
\end{tabular}
\end{table*}

In the following, the Pal1-I abundances will be always referred to our own solar
abundances. The measured Pal1-I abundances with respect to solar are also
reported in Table~\ref{abun}, as well as the dispersion ($\sigma$) of the
abundances. It is common practice to report, instead of $\sigma$, the standard
deviation of the mean of the abundances, i.e. the $\sigma$ values divided by the
square root of the number of lines used ($\sigma$/$\sqrt lines$). However, this
is a safe procedure only by assuming that each line provides an independent
measure of the element abundance and we prefer to report, instead, the $\sigma$
values obtained. 

We measure for Pal\,1-I an Iron content of [Fe/H]=-0.49$\pm$0.18. This is
perfectly compatible with the measures of R98b ([Fe/H]=-0.71$\pm$0.20).

\begin{table*}
\caption{Expected errors in the Pal1-I abundances due to estimated uncertainties
in the atmospheric parameters.}

\label{errors} 

\centering{}\begin{tabular}{ll|rr|rr|rr}
\hline 
Element  & Ion  & \multicolumn{2}{c|}{$\Delta$ \teff$=\pm$ 100 K} & \multicolumn{2}{c|}{$\Delta\log g=\pm0.50$} & \multicolumn{2}{c}{$\Delta\xi=\pm0.10$\,\kms}\tabularnewline
\hline 
Al  & I  & +0.06  & -0.06  & -0.01  & +0.01  & 0.00  & +0.01 \tabularnewline
Ba  & II  & +0.03  & -0.03  & +0.07  & -0.10  & -0.08  & +0.07 \tabularnewline
Ca  & I  & +0.10  & -0.10  & -0.06  & +0.04  & -0.04  & +0.04 \tabularnewline
Co  & I  & +0.12  & -0.11  & +0.03  & -0.02  & 0.00  & 0.00 \tabularnewline
Cr  & I  & +0.09  & -0.09  & -0.01  & +0.02  & -0.01  & +0.01 \tabularnewline
Fe  & I  & +0.08  & -0.08  & 0.00  & 0.00  & -0.03  & +0.03 \tabularnewline
Fe  & II  & -0.06  & +0.06  & +0.23  & -0.23  & -0.03  & +0.03 \tabularnewline
La  & II  & +0.02  & -0.01  & +0.22  & -0.21  & -0.01  & +0.02 \tabularnewline
Mg  & I  & +0.06  & -0.05  & -0.01  & +0.01  & -0.01  & +0.02 \tabularnewline
Mn  & I  & +0.09  & -0.09  & +0.04  & 0.00  & -0.01  & +0.03 \tabularnewline
Na  & I  & +0.07  & -0.08  & -0.11  & +0.08  & -0.03  & +0.02 \tabularnewline
Ni  & I  & +0.08  & -0.08  & +0.03  & -0.03  & -0.03  & +0.03 \tabularnewline
Sc  & II  & -0.01  & +0.01  & +0.07  & -0.22  & -0.03  & +0.03 \tabularnewline
Si  & I  & +0.02  & -0.01  & +0.06  & -0.04  & -0.01  & +0.01 \tabularnewline
Ti  & I  & +0.13  & -0.13  & -0.01  & +0.02  & -0.01  & +0.02 \tabularnewline
V  & I  & +0.15  & -0.15  & -0.01  & +0.01  & -0.02  & +0.02 \tabularnewline
Y  & II  & -0.01  & +0.02  & +0.21  & -0.21  & -0.05  & +0.06 \tabularnewline
Zn  & I  & -0.04  & +0.05  & +0.14  & -0.13  & -0.02  & +0.02 \tabularnewline
\hline
\end{tabular}
\end{table*}

We report in Table~\ref{errors} the changes in the abundances expected for 
variations in the atmospheric parameters of ($\Delta$ \teff$ = \pm$ 100 K$;
\Delta \log g = \pm 0.50 $; $\Delta \xi = \pm 0.10 $\,\kms), i.e. compatible
with their estimated uncertainty.

\section{Discussion}

In the ideal case, we would like to compare the Pal~1 abundance pattern to the
outcome of a self-consistent evolutionary model. Existing models however are not
yet able to return abundance ratios as detailed as the ones we can measure
observationally \citep[see, e.g., a review in][]{gnedin10}.  To gain some
insight into the possible origin of this cluster, we thus follow the common
practice of comparing its abundance ratios to those of other stellar systems.
This is further justified because abundances might be inherited, at least
partially, from the pre-enriched interstellar medium of a host system. Because
both Galactic and extra-galactic origins were postulated in the past for Pal~1,
these two possibilities are examined separately in the next sections.

\subsection{Pal\,1 as a Galactic cluster}

The iron abundance of Pal\,1 is typical of the thick disk, whose distribution
peaks around --0.7 \citep{ivezic}, and it is on the tail of the distribution of
the thin disk, which peaks around --0.15 \citep{holmberg}. This metallicity is
among the highest found in globular clusters and among the lowest found in open
clusters. Thus, from the point of view of its iron content, it is difficult to
classify Pal\,1.

\begin{figure}

  \includegraphics[width=1\columnwidth]{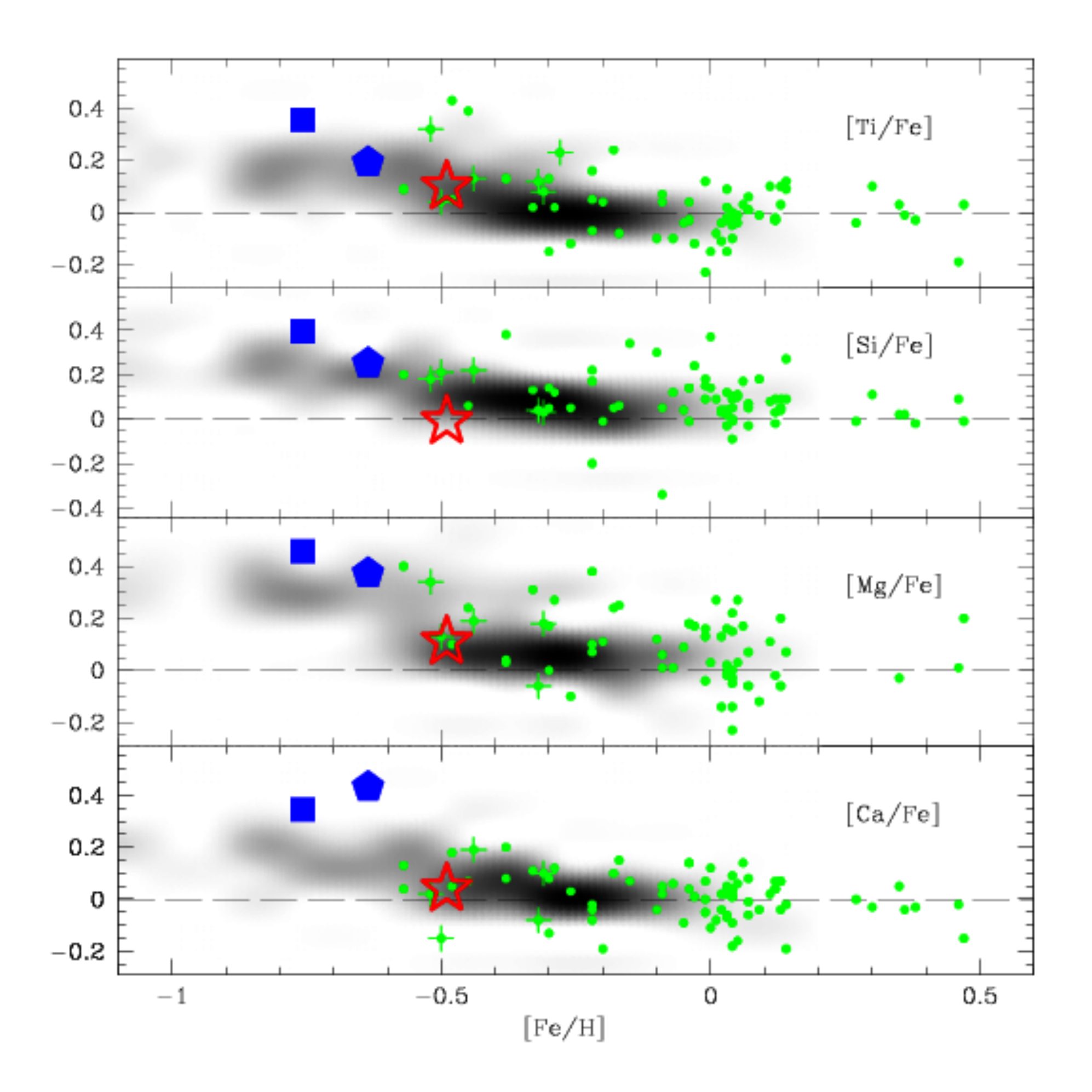}

  \caption{Pal1-I (big empty red star) $\alpha$-element abundance ratios are
    presented together with thick and thin Galactic disk stars (shaded area)
    from \citet[][]{reddy06,reddy03}. Also plotted are the Galactic OCs from
    \citet[][]{pancino10} and the literature compilation presented in that study
    (filled green circles) and the GCs 47~Tuc (big filled blue squares) and M~71
    (big filled pentagon) from \citet{koch08}, \citet[][]{ramirez02}. The OCs
    Be\,22, Be\,29 and Tom\,2 are also marked by a plus symbol.}\label{alphag}

\end{figure}

The ratio of $\alpha$ elements to iron (Fig.\,\ref{alphag}) is essentially
solar. From this point of view Pal\,1 is like thin disk stars and Open Clusters,
but markedly different from thick disk stars, which  show enhanced $\alpha$ to
iron ratios \citep[][and references therein]{Fuhrmann}. Note however, that from
the data of \citet{Fuhrmann}, at the metallicity  of Pal\,1 the thin disk
already displays  an $\alpha$ element enhancement, [Mg/Fe]$\sim +0.2$.

\begin{figure}
  \includegraphics[width=1\columnwidth]{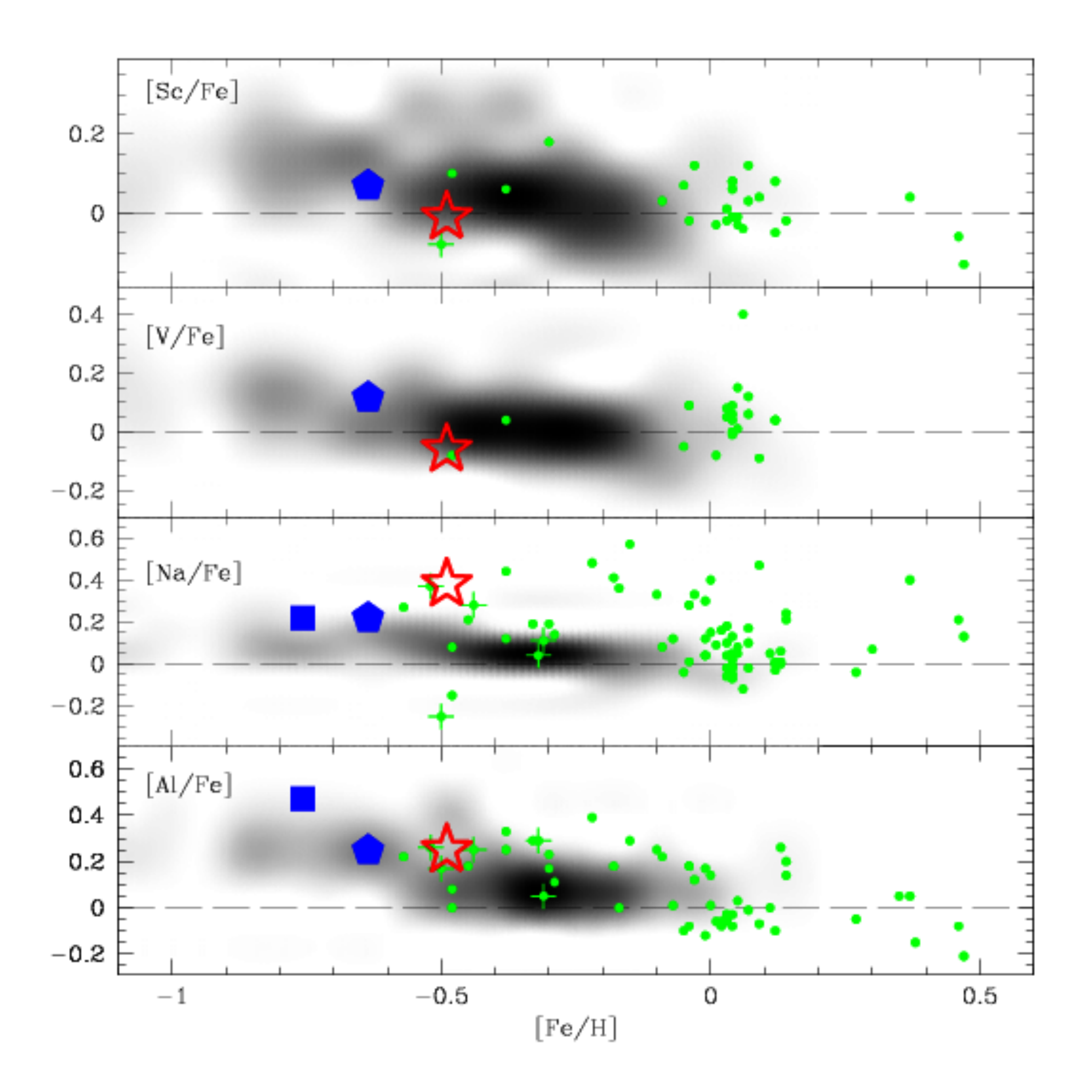}

  \caption{Pal1-I (big empty red star) light odd elements abundance ratios.
    Other symbols are the same as in Fig.~\ref{alphag}.}\label{lightg}

\end{figure}

The light odd elements (Na and Al, see Fig.\,\ref{lightg}), that are usually
associated to nucleosynthesis by proton captures, both show a moderate
enhancement. Both \citet{bensby05} and \citet{reddy06} noted that in disk stars 
Na and Al have opposite trends with metallicity, while Al behaves like $\alpha$
elements becoming enhanced with decreasing metallicity, Na is flat around
[Na/Fe]=+0.1. \citet{bensby05}  even claim a significative difference between 
thin and thick disk, the thin disk being slightly more Na enhanced. The
[Na/Fe]=+0.38 found by our analysis is  at odds, by a factor of 2 with the mild
Na enhancement found in disk stars. The [Al/Fe]=+0.25 is, instead perfectly in
line with the increase of this ratio with decreasing metallicity. Incidentally 
this puts Pal\,1 exactly between the thin and thick disk,  at this metallicity
(see Fig.\,8 of \citealt{bensby05}). One should question if this overabundance
of Na is real or if it is linked to differential non-local thermodynamic
equilibrium (NLTE) effects between  this giant in Pal\,1  and the dwarf and
sub-giant stars analyzed by \citet{bensby05} and \citet{reddy03,reddy06}.
According to the NLTE computations of \citet{gratton} a corrections of the order
of $\simeq$-0.1\,dex should be applied to both our Pal\,1 and solar abundances,
leaving, hence, the [Na/Fe] abundance ratio practically unchanged. The Na
abundances in disk dwarfs and subgiants were derived by  \citet{reddy03,reddy06}
using the 615\,nm, 616\,nm lines while \citet{bensby05} used also the same lines
adopted by us. In both these cases the NLTE corrections have been considered to
be small and were not included in the analysis. It thus appears unlikely that
the high Na abundance we derive, is due to differential NLTE effects between
dwarfs and giants. 

Another thing to be considered is whether the high Na abundance is indeed a
chemical signature of Pal\,1 or if it has to be understood in terms of the Na-O
anti-correlation found in Globular Clusters \citep[][and references
therein]{Carretta}. In Globular Clusters the ``unpolluted'' stars have
[O/Fe]$\sim +0.5$ and [Na/Fe]$\sim -0.3$ (see Fig.\,6 of \citealt{Carretta}), a
``polluted'' star, with [Na/Fe]=+0.4 can have an oxygen to iron ratio which is
1\,dex below this. Given that in Pal\,1-I the $\alpha$ to iron ratios are solar,
we expect [O/Fe] to be significantly sub-solar. However this possibility cannot
be checked, because the only measurable oxygen  line -- [O\,I] 630\,nm -- is in
our spectra superimposed to a telluric line and is, therefore, not suited for
abundance analysis.

The other odd light elements measured, Sc and V, may have a production channel 
by proton captures,  but may be also produced in nuclear statistical 
equilibrium, like other iron-peak elements. In Pal\,1 their ratio to iron
appears solar, like it is in  disk stars.

\begin{figure}
  \includegraphics[width=1\columnwidth]{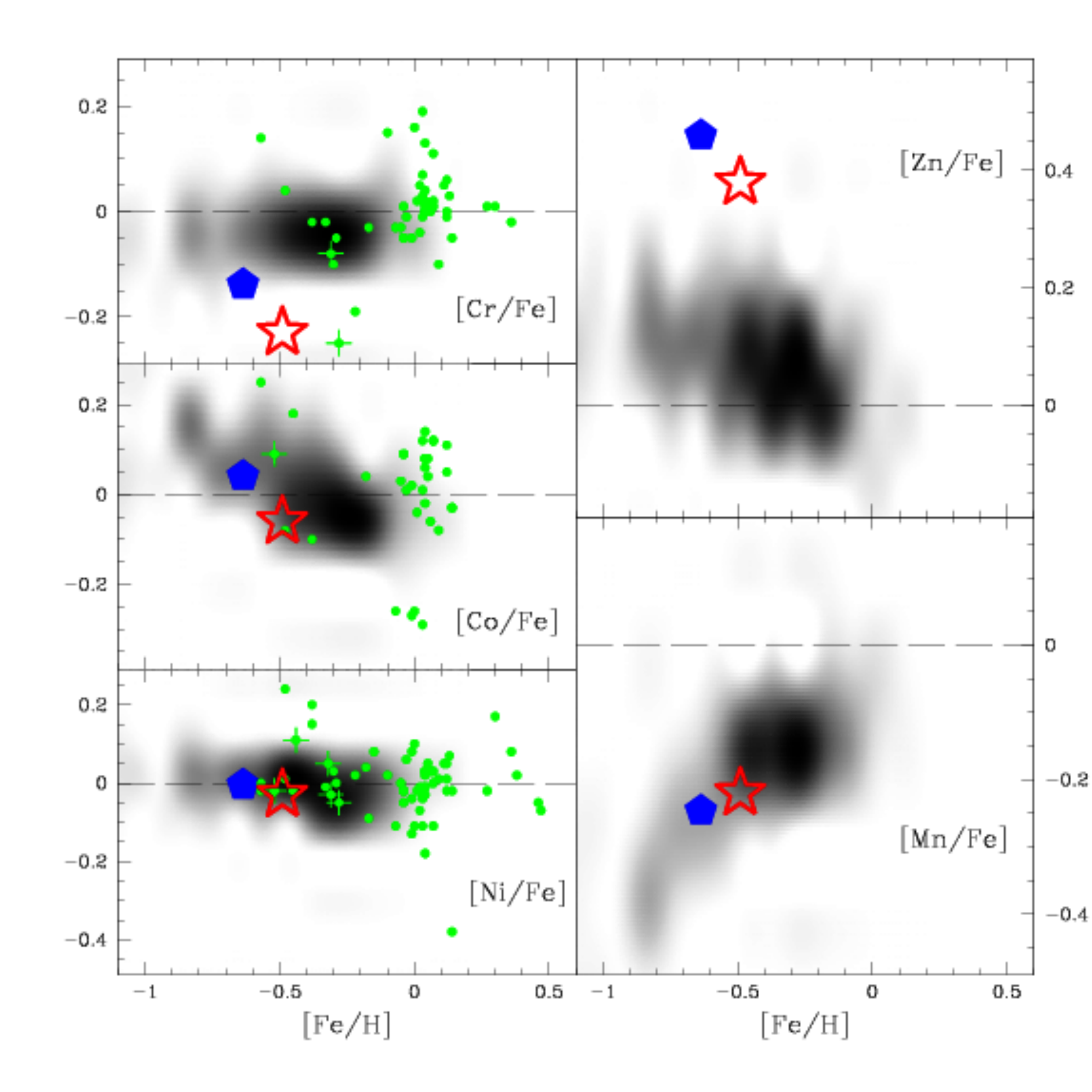}

  \caption{Pal1-I (big empty red star) abundance ratios for Fe-peak
    elements. Other symbols are the same as in
    Fig.~\ref{alphag}.}\label{fepeakg}

\end{figure}

Among the other iron-peak elements (Fig.\,\ref{fepeakg}) what stands out are
the  underabundance of Cr and the overabundance of Zn. For Cr one should be
aware of the possible NLTE effects, as are found in metal-poor stars
\citep{bonifacio09,bergemann}. For Zn, according to the computations  of
\citet{Takeda} the NLTE correction should be only of the order of -0.1\,dex. It
is anyway remarkable that for both these elements the ratio to iron appears
similar to what found in the Globular Cluster M71 \citep{ramirez02}. Therefore
it is likely that these ratios are specific signatures of Pal\,1.

\begin{figure}
  \includegraphics[width=1\columnwidth]{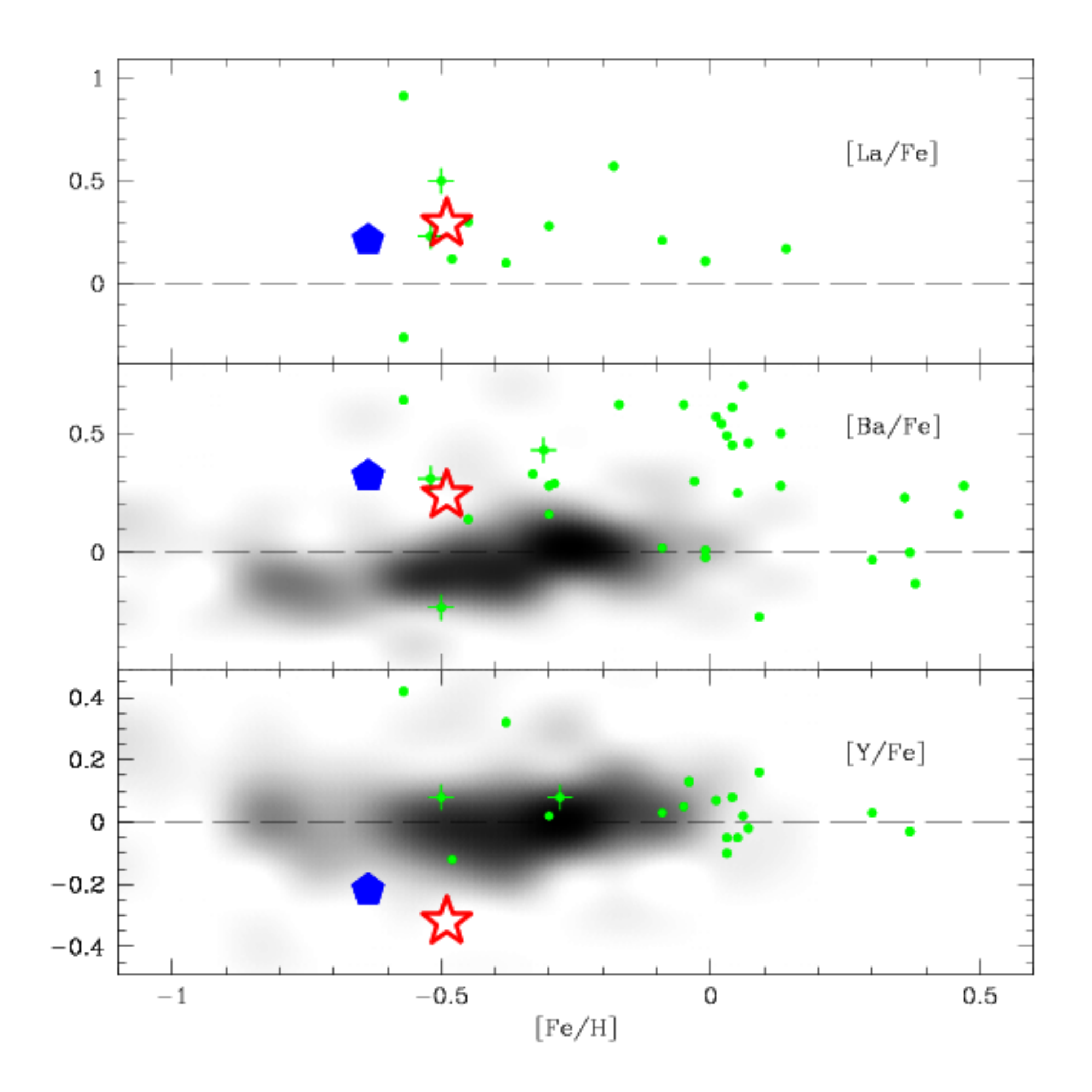}

  \caption{Pal1-I (big empty red star) heavy element abundance ratios. Other
    symbols are the same as in Fig.~\ref{alphag}.}\label{heavyg}

\end{figure}

For neutron capture elements (Fig.\,\ref{heavyg}) the most relevant feature,
with respect to disk stars of similar metallicities,  is the underabundance of Y
and overabundance of Ba. It is tempting to interpret this as an overabundance of
second-peak $s-$process elements (Ba), to light $s-$process elements (Y) as can
be expected when the $s-$process is dominated by moderately metal-poor stars
\citep[see][and references therein]{Busso}.

\subsection{The possible extra-galactic origin of Pal\,1}

Because Pal~1 is peculiar in terms of age and metallicity compared to other GCs,
a number of studies tried to associate it to different Galactic sub-structures.
In particular, \citet[][]{frinchaboy04} proposed the GASS, and
\citet[][]{forbes10} proposed the putative CMa dwarf \citep[see
also][]{martin04}.  In view of the ambiguous nature of these two structures, we
add also the Sagittarius dSph galaxy to the set of template extra-galactic
systems: it is the only galaxy with measured abundance ratios, and whose
metallicity  reaches values as high as that of Pal~1
\citep[][]{monaco05,sbordone07}.

\begin{figure}

  \includegraphics[width=1\columnwidth]{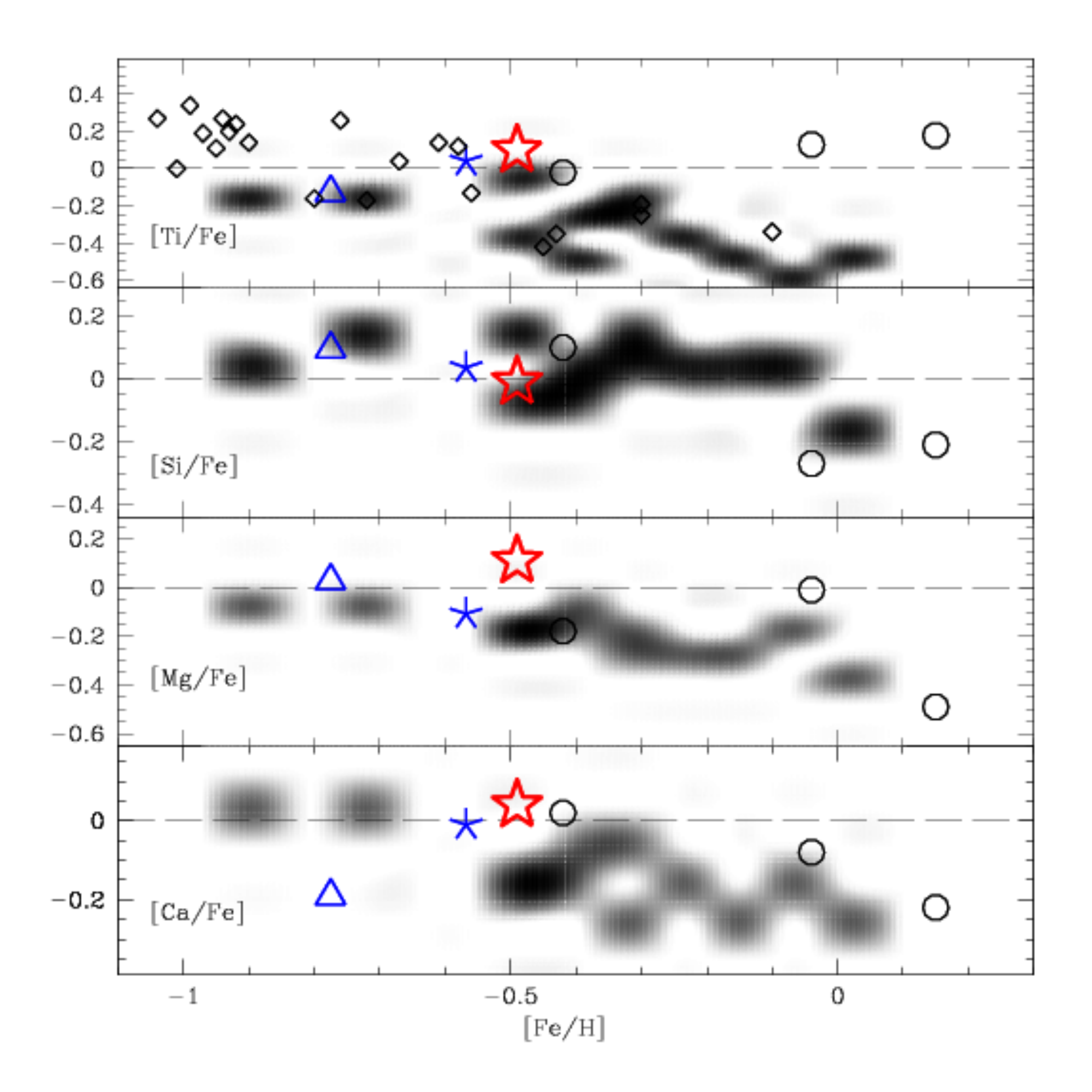}

  \caption{Pal1-I (big empty red star) $\alpha$-element abundance ratios are
    presented together with Sagittarius dSph stars (grey shaded area) from S07,
    and stars in the GCs Pal~12 (big open blue triangle) and Ter~7 (big blue
    asterisk) from \citet[][]{cohen04} and S07, respectively. Stars in the CMa
    overdensity region in the background of the OC NGC\,2477 are plotted as big
    open circles \citep[][]{sbordone05}. M-giants in the GASS from
    \citet[][]{chou10} are plotted as open diamonds.}\label{alphax}

\end{figure}

\begin{figure}
  \includegraphics[width=1\columnwidth]{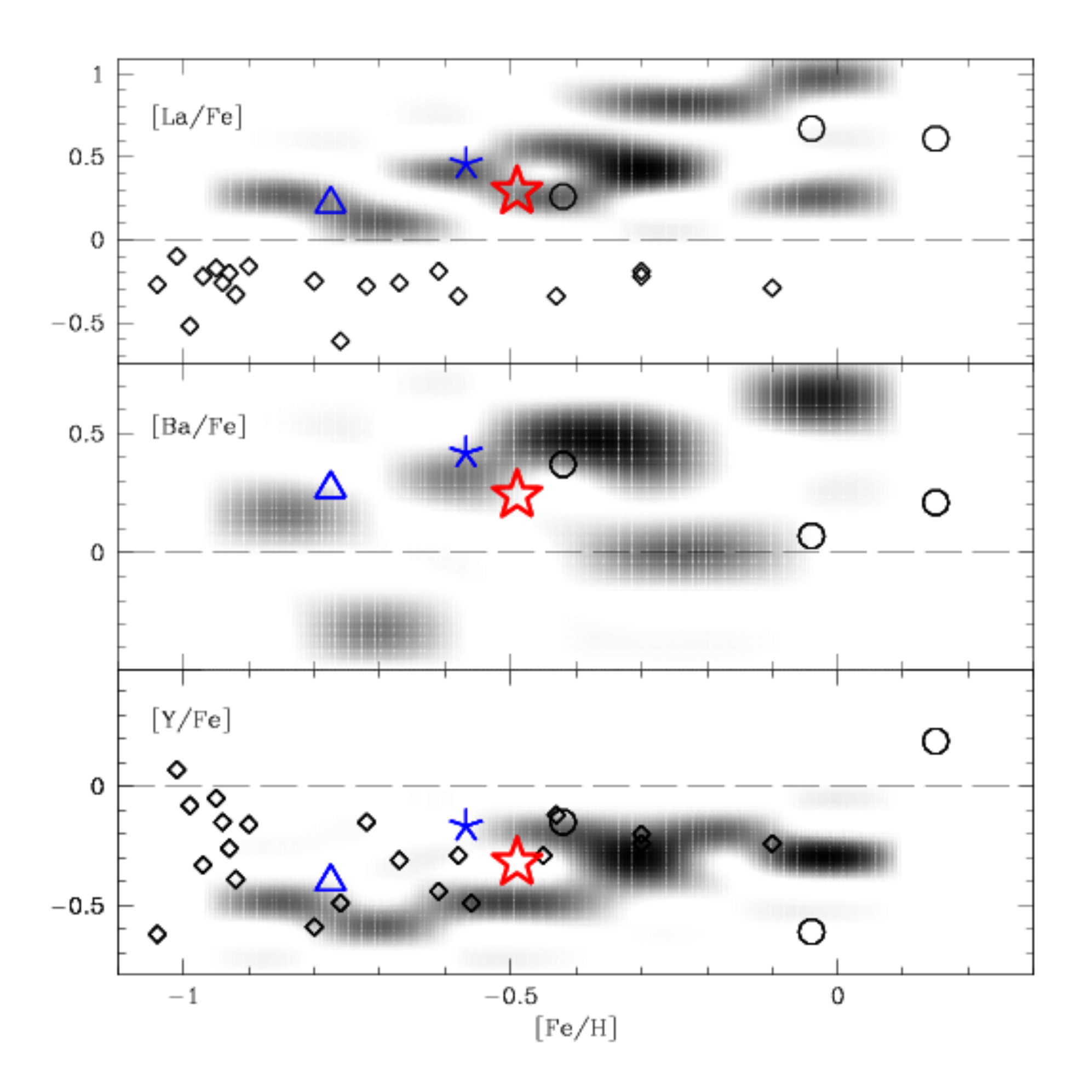}

  \caption{Pal1-I (big empty red star) heavy element abundance ratios. Other
    symbols are the same as in Fig.~\ref{alphax}.}\label{heavyx}

\end{figure}

\begin{figure}
  \includegraphics[width=1\columnwidth]{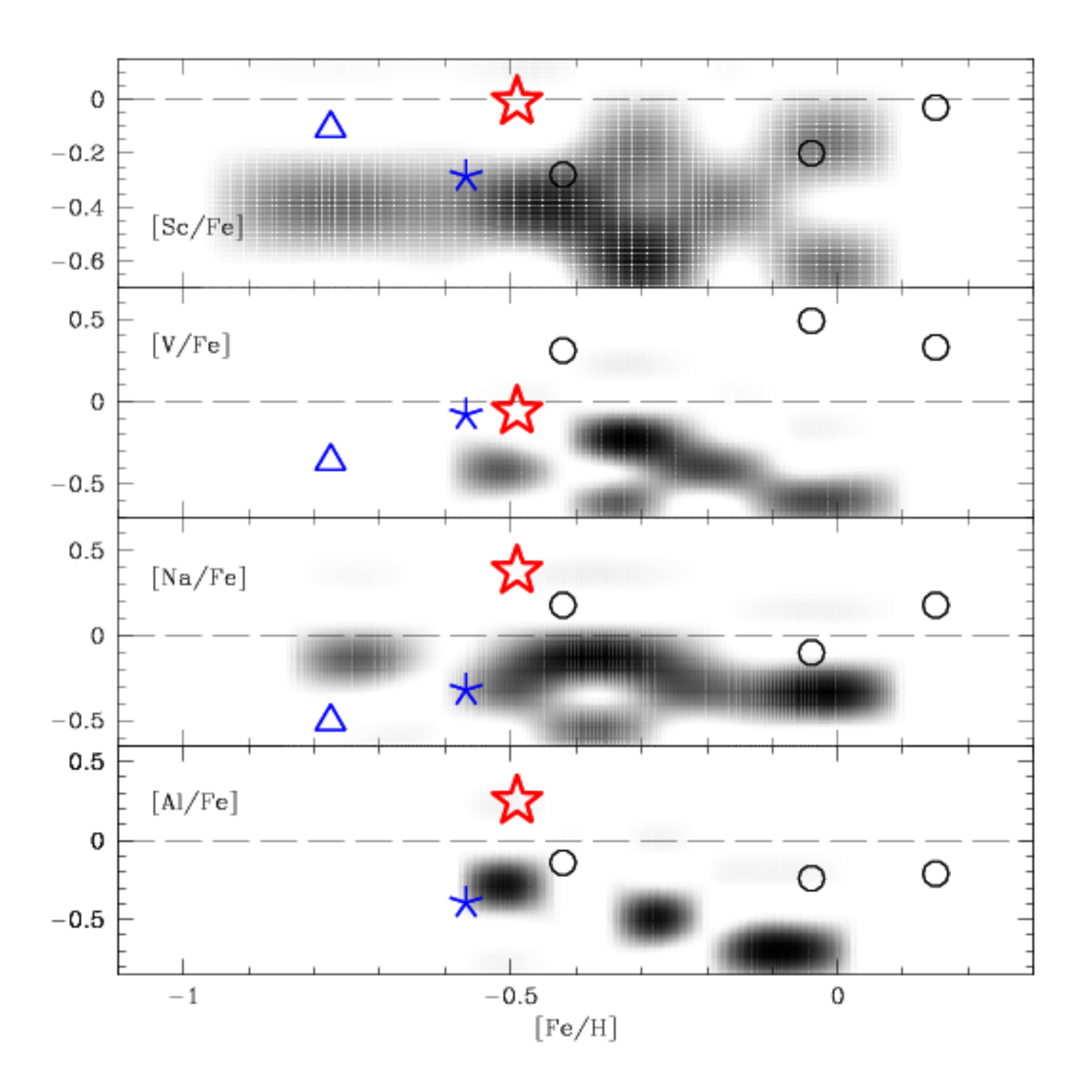}

  \caption{Pal1-I (big empty red star) light odd elements abundance ratios.
    Other symbols are the same as in Fig.~\ref{alphax}.}\label{lightx}

\end{figure}

Therefore in Figs.\,\ref{alphax} to \ref{lightx} we show again the abundance
ratios of Pal\,1-I, but this time compared  to Sagittarius dSph stars from
\citet[][hereafter S07]{sbordone07},  stars in the GCs Pal~12  and Ter~7 
(likely associated to Sgr, \citealt{cohen04} and S07), stars in the CMa
overdensity region  \citep[][]{sbordone05} and stars in the GASS
\citep[][]{chou10}. 

\citet[][]{chou10} have recently, presented Ti, La and Y abundances for 21
M-giants in the GASS (open diamonds in Figs.\,\ref{alphax} and \ref{heavyx}).
Pal\,1 presents Ti and Y content compatible with them, while the La abundance is
remarkably different. Detailed chemical abundances for more stars in Pal\,1 and
for additional elements for stars in the GASS are required to reliably assess
the compatibility of the two chemical patterns. At the present stage, we can
only conclude that the association of Pal\,1 and the GASS is not supported by
our abundance analysis. The AMR of the GASS was derived by \cite{frinchaboy04}
using both GCs and OCs belonging to the area of interest. Some of the OCs they
considered, namely Be\,22, Be\,29 and Tom\,2, have measured abundances and are
included in Figs.\,\ref{alphag}---\ref{heavyg} (plus symbols). These OCs have
abundance patterns qualitatively not dissimilar with respect to the rest of the
OCs plotted in the figures, which are also similar to Pal~1. 

Data are definitely more sparse for the case of CMa. To date only  three giant
stars  in the background of the OC NGC\,2477 were studied \citep[][hereafter
S05, big open circles in the Figs.\,\ref{alphax}---\ref{heavyx}]{sbordone05}.
S05 suggested that the CMa structure should have undergone a level of chemical
processing compatible with that of the Galactic disk. Because Pal\,1 chemical
properties are also compatible with those of the disk, an association to CMa
might be possible. Indeed, the most metal-poor among the stars studied in S05
has an iron content similar to Pal\,1-I ({[}Fe/H{]}=-0.42) and the chemical
composition of the two objects is also quite similar, particularly in the
$\alpha$ elements and neutron capture elements (see
Figs.\,\ref{alphax}---\ref{heavyx}). 

\begin{figure}
  \includegraphics[width=1\columnwidth]{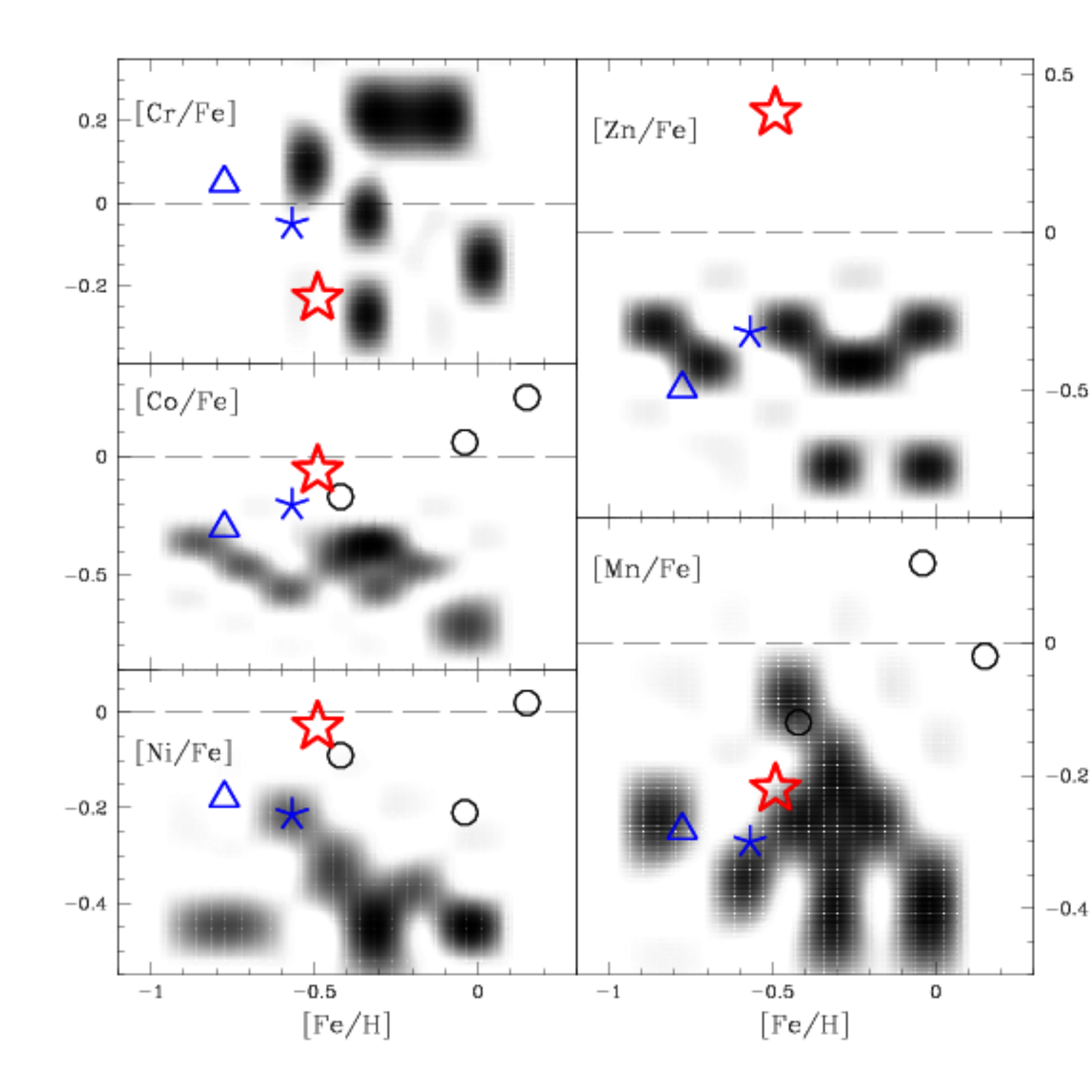}

  \caption{Pal1-I (big empty red star) abundance ratios for Fe-peak
    elements. Other symbols are the same as in
    Fig.~\ref{alphax}.}\label{fepeakx}

\end{figure}

We also compare with the Sgr dSph, which, as mentioned above, is the only among
MW satellite to have a mean metallicity similar to that of Pal\,1
\citep[][]{monaco02,monaco05}. Sgr stars studied by \citet[][hereafter
S07]{sbordone07} are plotted in Figs.\,\ref{alphax}---\ref{heavyx} as a gray
shaded area. The GCs Ter\,7 (S07, big  asterisk) and Pal\,12 \citep[][big open 
triangle]{cohen04} are also plotted since they are believed to belong to the Sgr
GCs system, either in the main body (Ter\,7) or in the Sgr trailing tail
(Pal\,12). There are some remarkable similarities of Pal\,1 abundances with
Ter~7 for $\alpha$ elements and neutron capture elements
(Figs.\,\ref{alphax}---\ref{heavyx}). For the latter there is a similarity also
with the Sgr field stars. The sodium overabundance seems however a distinctive
feature of Pal\,1-I (Fig.\,\ref{lightx}). Additionally, Pal\,1 does not show the
peculiar Sgr Iron-peak and light odd elements (Figs.\,\ref{fepeakx} and
\ref{lightx}) abundance ratios. 

\section{Conclusion}

We have presented the first high resolution chemical abundances of 18 among
$\alpha$, Iron-peak, light and heavy elements for one giant in the peculiar,
young cluster Palomar\,1. Pal\,1 chemical abundance pattern is similar to that
of the Galactic OCs population and significantly different from that of stars in
the Sgr dwarf spheroidal galaxy.  The different chemical composition with
respect to stars analyzed in the GASS, would argue against the associations of
Pal\,1 with this structure. More data is, however, needed to reliably assess the
compatibility of the two systems chemical patterns. If the association between
Pal\,1 and CMa will be proved, this would imply that the CMa overdensity has
undergone a degree of chemical processing similar to the Galactic disk, in
agreement with \cite{sbordone05}. 

Pal\,1 might then be a GC which experienced a peculiar chemical evolution or an
OC ejected from the Galactic disk, which high concentration and flat MF would be
the result of its peculiar dynamical evolution. Only a reconstruction of the
cluster's orbit would permit discriminating between the two possibilities. In
this respect we remark that, as pointed out by \cite{penarubbia05}, radial
velocities and positions are not sufficient to claim a common orbit;
measurements of proper motions are needed to break degeneracies. A measure of
the Pal\,1 proper motion is therefore crucial to disentangle the origin of this
peculiar system.

\begin{acknowledgements} 

We would like to thank Elena Pancino for providing the literature open clusters
data used in Figs.\,\ref{alphag}---\ref{heavyg} in electronic format and for
useful discussion about the DAOSPEC code.

\end{acknowledgements}

\Online

\appendix 

\section{Individual line data}\label{appendix} 

The following table reports the line list and atomic parameters adopted for
Pal1-I and the Sun. For the Mn and Co lines we adopted the HFS of
\citet[][]{prochaska00}. The measured equivalent width and the corresponding
abundance obtained for each line are also reported.

\begin{table*} 
\caption{Adopted line list and atomic parameters. The measured equivalent width and the
corresponding abundances obtained for each line are also reported for both Pal1-I
and the Sun.} 
\label{lines} 
\begin{center} 
\begin{tabular}{llrrrrrrr} 
\hline 
Element & Ion  & $\lambda$ & log gf  & $\chi$      & EW       &$\epsilon$ & EW       & $\epsilon$\\
        &      & ($\AA$)   &         & eV          & (m$\AA$) &           & (m$\AA$) &           \\
        &      &           &         &             & \object{Pal1-I}         &          & \object{Sun}\\ 
\hline
 Al & I & 6696.023 & -1.347 &3.143 & 42.0 & 5.99 & 38.5 &6.25 \\ 
 Al & I & 6698.673 & -1.647 &3.143 & 26.0 & 5.97 & 21.1 &6.20 \\ 
 Ba &II & 5853.668 & -1.000 &0.604 & 94.5 & 2.01 & 66.9 &2.30 \\ 
 Ba &II & 6496.897 & -0.377 &0.604 &139.2 & 1.98 &104.2 &2.20 \\ 
 Ca & I & 5581.965 & -0.555 &2.523 & 88.4 & 5.88 & 97.4 &6.39 \\ 
 Ca & I & 5601.277 & -0.523 &2.526 & 89.1 & 5.86 &111.8 &6.58 \\ 
 Ca & I & 5857.451 &  0.240 &2.933 &115.6 & 5.92 &133.8 &6.19 \\ 
 Ca & I & 6455.598 & -1.340 &2.523 & 51.1 & 5.75 &----- &----- \\ 
 Ca & I & 6471.662 & -0.686 &2.526 & 97.3 & 6.12 & 94.8 &6.42 \\ 
 Ca & I & 6493.781 & -0.109 &2.521 &123.7 & 6.07 &129.3 &6.33 \\ 
 Ca & I & 6572.779 & -4.240 &0.000 & 55.1 & 5.78 &----- &----- \\ 
 Ca & I & 6717.681 & -0.524 &2.709 &100.5 & 6.07 &117.4 &6.38 \\ 
 Co & I & 6814.942 & HFS & HFS & 37.5 & 4.42 & 19.6 &4.97 \\ 
 Cr & I & 5783.063 & -0.500 &3.323 & 21.1 & 4.95 & 32.6 &5.69 \\ 
 Cr & I & 5783.850 & -0.295 &3.322 & 41.5 & 5.23 & 45.0 &5.72 \\ 
 Cr & I & 5784.969 & -0.380 &3.321 & 26.0 & 4.96 &----- &----- \\ 
 Cr & I & 5787.918 & -0.083 &3.322 & 33.5 & 4.84 & 47.2 &5.55 \\ 
 Cr & I & 6330.091 & -2.920 &0.941 & 23.5 & 4.69 & 25.9 &5.63 \\ 
 Fe & I & 5525.539 & -1.084 &4.230 & 64.1 & 7.13 & 55.2 &7.35 \\ 
 Fe & I & 5549.948 & -2.910 &3.694 & 21.9 & 7.37 &  8.4 &7.46 \\ 
 Fe & I & 5554.882 & -0.440 &4.548 & 77.8 & 7.15 & 94.7 &7.57 \\ 
 Fe & I & 5560.207 & -1.190 &4.434 & 43.7 & 6.99 & 51.4 &7.55 \\ 
 Fe & I & 5567.391 & -2.564 &2.608 & 77.6 & 7.10 & 68.9 &7.61 \\ 
 Fe & I & 5587.574 & -1.850 &4.143 & 30.9 & 7.04 &----- &----- \\ 
 Fe & I & 5600.224 & -1.420 &4.260 & 36.0 & 6.86 &----- &----- \\ 
 Fe & I & 5619.587 & -1.700 &4.386 & 27.2 & 7.06 & 34.2 &7.69 \\ 
 Fe & I & 5624.022 & -1.480 &4.386 & 38.2 & 7.10 & 54.5 &7.87 \\ 
 Fe & I & 5633.946 & -0.270 &4.991 & 53.7 & 6.90 & 70.3 &7.53 \\ 
 Fe & I & 5635.822 & -1.890 &4.256 & 29.1 & 7.15 & 35.4 &7.77 \\ 
 Fe & I & 5636.696 & -2.610 &3.640 & 16.1 & 6.82 & 19.3 &7.54 \\ 
 Fe & I & 5638.262 & -0.870 &4.220 & 76.7 & 7.19 & 78.6 &7.54 \\ 
 Fe & I & 5641.434 & -1.180 &4.256 & 65.1 & 7.27 & 66.7 &7.68 \\ 
 Fe & I & 5649.987 & -0.920 &5.099 & 27.3 & 7.07 & 35.7 &7.62 \\ 
 Fe & I & 5650.706 & -0.960 &5.085 & 35.0 & 7.28 &----- &----- \\ 
 Fe & I & 5652.318 & -1.950 &4.260 & 32.2 & 7.29 & 24.6 &7.61 \\ 
 Fe & I & 5653.865 & -1.640 &4.386 & 37.2 & 7.23 & 38.9 &7.71 \\ 
 Fe & I & 5655.493 & -0.796 &4.260 & 60.0 & 6.80 & 73.3 &7.54 \\ 
 Fe & I & 5667.518 & -1.576 &4.178 & 43.3 & 7.08 & 51.7 &7.77 \\ 
 Fe & I & 5679.023 & -0.920 &4.652 & 46.7 & 7.02 &----- &----- \\ 
 Fe & I & 5680.240 & -2.580 &4.186 & 16.7 & 7.42 &----- &----- \\ 
 Fe & I & 5691.497 & -1.520 &4.301 & 34.7 & 6.96 & 41.5 &7.59 \\ 
 Fe & I & 5701.544 & -2.216 &2.559 & 79.6 & 6.72 & 85.7 &7.53 \\ 
 Fe & I & 5705.464 & -1.355 &4.301 & 35.9 & 6.82 & 38.3 &7.36 \\ 
 Fe & I & 5717.833 & -1.130 &4.284 & 56.3 & 7.05 &----- &----- \\ 
 Fe & I & 5731.762 & -1.300 &4.256 & 56.7 & 7.19 & 58.7 &7.65 \\ 
 Fe & I & 5732.275 & -1.560 &4.991 & 13.5 & 7.17 & 14.1 &7.59 \\ 
 Fe & I & 5752.032 & -1.177 &4.549 & 49.1 & 7.23 & 56.0 &7.81 \\ 
 Fe & I & 5753.121 & -0.688 &4.260 & 80.6 & 7.15 & 82.6 &7.45 \\ 
 Fe & I & 5760.344 & -2.490 &3.642 & 29.3 & 7.07 & 22.8 &7.52 \\ 
 Fe & I & 5775.081 & -1.298 &4.220 & 50.7 & 7.02 &----- &----- \\ 
 Fe & I & 5778.453 & -3.430 &2.588 & 39.6 & 7.03 & 21.5 &7.39 \\ 
 Fe & I & 5784.657 & -2.532 &3.396 & 26.6 & 6.76 &----- &----- \\ 
 Fe & I & 5793.913 & -1.700 &4.220 & 22.6 & 6.75 & 35.2 &7.56 \\ 
 Fe & I & 5806.717 & -1.050 &4.607 & 44.4 & 7.05 & 55.9 &7.64 \\ 
 Fe & I & 5814.805 & -1.970 &4.283 & 31.2 & 7.30 & 22.8 &7.60 \\ 
 Fe & I & 5816.373 & -0.601 &4.548 & 78.5 & 7.35 &----- &----- \\ 
 Fe & I & 5835.098 & -2.370 &4.256 & 17.7 & 7.31 & 14.4 &7.71 \\ 
 Fe & I & 5838.370 & -2.340 &3.943 & 23.9 & 7.11 & 21.2 &7.59 \\ 
 Fe & I & 5852.217 & -1.330 &4.548 & 31.4 & 6.96 & 40.9 &7.60 \\ 
 Fe & I & 5859.586 & -0.419 &4.549 & 65.0 & 6.84 & 75.4 &7.37 \\ 
 Fe & I & 5862.357 & -0.127 &4.549 & 71.9 & 6.72 & 88.7 &7.34 \\ 
 Fe & I & 5881.279 & -1.840 &4.607 & 17.4 & 7.16 & 15.5 &7.55 \\ 
 Fe & I & 5883.813 & -1.360 &3.960 & 64.7 & 7.09 & 75.7 &7.66 \\ 
 \hline 
\end{tabular} 
\end{center} 
\end{table*} 
\addtocounter{table}{-1} 
\begin{table*} 
\caption{Adopted line list and atomic parameters. The measured equivalent width and the
corresponding abundances obtained for each line are also reported for both Pal1-I
and the Sun (continued).} 
\begin{center} 
\begin{tabular}{llrrrrrrr} 
\hline 
Element & Ion  & $\lambda$ & log gf  & $\chi$      & EW       &$\epsilon$ & EW       & $\epsilon$\\
        &      & ($\AA$)   &         & eV          & (m$\AA$) &           & (m$\AA$) &           \\
        &      &           &         &             & \object{Pal1-I}         &          & \object{Sun}\\ 
\hline
Fe & I & 5905.671 & -0.730 &4.652 & 38.6 & 6.64 & 58.6 &7.40 \\ 
 Fe & I & 5909.970 & -2.587 &3.211 & 42.2 & 6.95 & 34.2 &7.44 \\ 
 Fe & I & 5916.247 & -2.994 &2.453 & 57.1 & 6.82 & 55.8 &7.58 \\ 
 Fe & I & 5927.786 & -1.090 &4.652 & 36.3 & 6.95 & 41.1 &7.46 \\ 
 Fe & I & 5929.667 & -1.410 &4.548 & 41.2 & 7.26 & 40.0 &7.66 \\ 
 Fe & I & 5952.716 & -1.440 &3.984 & 61.8 & 7.12 & 64.7 &7.58 \\ 
 Fe & I & 5976.777 & -1.243 &3.943 & 70.5 & 7.12 & 74.4 &7.67 \\ 
 Fe & I & 5984.815 & -0.196 &4.733 & 77.5 & 7.12 & 86.5 &7.52 \\ 
 Fe & I & 6003.010 & -1.120 &3.881 & 83.8 & 7.19 & 87.1 &7.52 \\ 
 Fe & I & 6007.960 & -0.597 &4.652 & 55.7 & 6.91 & 64.6 &7.48 \\ 
 Fe & I & 6012.206 & -4.038 &2.223 & 25.7 & 6.89 & 25.0 &7.72 \\ 
 Fe & I & 6015.243 & -4.680 &2.223 & 17.0 & 7.29 &----- &----- \\ 
 Fe & I & 6024.049 & -0.120 &4.548 & 93.7 & 7.12 &110.9 &7.46 \\ 
 Fe & I & 6027.051 & -1.089 &4.076 & 65.8 & 7.01 &----- &----- \\ 
 Fe & I & 6055.992 & -0.460 &4.733 & 61.9 & 6.98 & 75.2 &7.46 \\ 
 Fe & I & 6078.491 & -0.321 &4.796 & 66.8 & 7.05 & 79.5 &7.59 \\ 
 Fe & I & 6078.999 & -1.120 &4.652 & 48.1 & 7.24 & 47.2 &7.59 \\ 
 Fe & I & 6082.708 & -3.573 &2.223 & 48.8 & 6.93 & 36.3 &7.51 \\ 
 Fe & I & 6085.258 & -3.095 &2.758 & 59.3 & 7.31 & 42.8 &7.70 \\ 
 Fe & I & 6093.643 & -1.500 &4.607 & 31.7 & 7.20 & 30.4 &7.60 \\ 
 Fe & I & 6096.662 & -1.930 &3.984 & 33.9 & 6.98 & 37.9 &7.58 \\ 
 Fe & I & 6098.244 & -1.880 &4.558 & 22.9 & 7.30 & 15.6 &7.54 \\ 
 Fe & I & 6102.171 & -0.516 &4.835 & 67.3 & 7.30 & 82.4 &7.85 \\ 
 Fe & I & 6187.987 & -1.720 &3.943 & 50.9 & 7.10 & 48.6 &7.54 \\ 
 Fe & I & 6200.313 & -2.437 &2.608 & 92.0 & 7.23 &----- &----- \\ 
 Fe & I & 6220.776 & -2.460 &3.881 & 29.1 & 7.28 & 20.2 &7.61 \\ 
 Fe & I & 6290.965 & -0.774 &4.733 & 54.1 & 7.12 & 67.2 &7.76 \\ 
 Fe & I & 6315.306 & -1.232 &4.143 & 70.0 & 7.31 & 70.6 &7.83 \\ 
 Fe & I & 6315.811 & -1.710 &4.076 & 35.0 & 6.89 & 42.3 &7.61 \\ 
 Fe & I & 6322.685 & -2.426 &2.588 & 82.2 & 6.95 & 77.7 &7.56 \\ 
 Fe & I & 6330.838 & -1.740 &4.733 & 30.7 & 7.54 & 32.8 &8.00 \\ 
 Fe & I & 6344.148 & -2.923 &2.433 & 81.1 & 7.24 & 67.7 &7.71 \\ 
 Fe & I & 6380.743 & -1.376 &4.186 & 53.2 & 7.09 & 52.1 &7.60 \\ 
 Fe & I & 6392.538 & -4.030 &2.279 & 34.0 & 7.11 & 16.8 &7.51 \\ 
 Fe & I & 6408.018 & -1.018 &3.686 & 89.5 & 6.99 &101.5 &7.59 \\ 
 Fe & I & 6436.406 & -2.460 &4.186 & 17.9 & 7.31 &  9.8 &7.52 \\ 
 Fe & I & 6469.192 & -0.770 &4.835 & 44.5 & 6.99 & 60.1 &7.61 \\ 
 Fe & I & 6475.624 & -2.942 &2.559 & 73.4 & 7.21 & 60.4 &7.69 \\ 
 Fe & I & 6496.466 & -0.570 &4.795 & 57.0 & 7.02 & 65.0 &7.45 \\ 
 Fe & I & 6533.928 & -1.460 &4.558 & 34.9 & 7.15 & 42.3 &7.72 \\ 
 Fe & I & 6569.209 & -0.420 &4.733 & 66.8 & 7.01 & 92.3 &7.65 \\ 
 Fe & I & 6575.016 & -2.710 &2.588 & 72.5 & 6.98 &----- &----- \\ 
 Fe & I & 6597.557 & -1.070 &4.795 & 42.6 & 7.20 & 43.0 &7.57 \\ 
 Fe & I & 6608.024 & -4.030 &2.279 & 37.3 & 7.17 & 16.1 &7.48 \\ 
 Fe & I & 6609.110 & -2.692 &2.559 & 80.6 & 7.11 & 66.9 &7.56 \\ 
 Fe & I & 6627.540 & -1.680 &4.548 & 30.9 & 7.27 & 27.6 &7.64 \\ 
 Fe & I & 6633.746 & -0.799 &4.558 & 56.9 & 6.97 & 70.3 &7.55 \\ 
 Fe & I & 6703.566 & -3.160 &2.758 & 57.0 & 7.27 & 37.7 &7.62 \\ 
 Fe & I & 6715.382 & -1.640 &4.607 & 22.7 & 7.09 & 28.2 &7.67 \\ 
 Fe & I & 6725.353 & -2.300 &4.103 & 25.0 & 7.24 & 16.5 &7.53 \\ 
 Fe & I & 6726.666 & -1.133 &4.607 & 59.3 & 7.43 & 47.3 &7.59 \\ 
 Fe & I & 6750.150 & -2.621 &2.424 & 90.1 & 7.10 & 75.6 &7.52 \\ 
 Fe & I & 6752.705 & -1.204 &4.638 & 34.0 & 6.96 & 36.5 &7.44 \\ 
 Fe & I & 6786.856 & -2.070 &4.191 & 27.9 & 7.18 & 26.4 &7.64 \\ 
 Fe & I & 6806.843 & -3.210 &2.727 & 43.6 & 6.99 & 35.9 &7.60 \\ 
 Fe & I & 6810.257 & -0.986 &4.607 & 30.4 & 6.62 & 51.7 &7.46 \\ 
 Fe & I & 6820.369 & -1.320 &4.638 & 45.5 & 7.32 & 41.6 &7.63 \\ 
 Fe & I & 6828.590 & -0.920 &4.638 & 65.9 & 7.37 & 58.1 &7.55 \\ 
 Fe & I & 6839.830 & -3.450 &2.559 & 43.5 & 7.03 & 32.4 &7.59 \\ 
 Fe & I & 6842.679 & -1.320 &4.638 & 42.2 & 7.25 & 41.0 &7.62 \\ 
 Fe & I & 6843.648 & -0.930 &4.548 & 54.8 & 7.03 & 61.8 &7.54 \\ 
 Fe & I & 6857.249 & -2.150 &4.076 & 16.8 & 6.82 & 20.1 &7.48 \\ 
 Fe & I & 6858.145 & -0.930 &4.607 & 44.4 & 6.88 & 50.6 &7.40 \\ 
 Fe &II & 5534.847 & -2.865 &3.245 & 74.0 & 7.21 & 61.4 &7.60 \\ 
 Fe &II & 5991.376 & -3.647 &3.153 & 47.1 & 7.20 & 31.6 &7.57 \\ 
 \hline 
\end{tabular} 
\end{center} 
\end{table*} 
\addtocounter{table}{-1} 
\begin{table*} 
\caption{Adopted line list and atomic parameters. The measured equivalent width and the
corresponding abundances obtained for each line are also reported for both Pal1-I
and the Sun (continued).} 
\begin{center} 
\begin{tabular}{llrrrrrrr} 
\hline 
Element & Ion  & $\lambda$ & log gf  & $\chi$      & EW       &$\epsilon$ & EW       & $\epsilon$\\
        &      & ($\AA$)   &         & eV          & (m$\AA$) &           & (m$\AA$) &           \\
        &      &           &         &             & \object{Pal1-I}         &          & \object{Sun}\\ 
\hline
 Fe &II & 6084.111 & -3.881 &3.199 & 35.3 & 7.19 & 20.0 &7.52 \\ 
 Fe &II & 6247.557 & -2.435 &3.892 & 49.3 & 6.86 & 54.8 &7.60 \\ 
 Fe &II & 6416.919 & -2.877 &3.892 & 43.3 & 7.15 & 40.1 &7.70 \\ 
 Fe &II & 6432.680 & -3.687 &2.891 & 55.8 & 7.14 & 42.1 &7.59 \\ 
 Fe &II & 6516.080 & -3.432 &2.891 & 52.4 & 6.80 & 56.1 &7.65 \\ 
 La &II & 6390.477 & -1.410 &0.321 & 22.3 & 0.97 &----- &----- \\ 
 La &II & 6774.268 & -1.708 &0.126 & 18.4 & 0.90 &----- &----- \\ 
 Mg & I & 5711.088 & -1.833 &4.346 & 92.0 & 7.20 &105.2 &7.67 \\ 
 Mg & I & 6318.717 & -1.730 &5.108 & 42.2 & 7.05 & 50.6 &7.44 \\ 
 Mg & I & 6319.237 & -1.950 &5.108 & 33.6 & 7.10 & 31.7 &7.34 \\ 
 Mn & I & 6013.513 & HFS & HFS & 83.6 & 4.63 & 88.0 &5.34 \\ 
 Na & I & 5682.633 & -0.700 &2.102 &103.5 & 6.15 &106.4 &6.23 \\ 
 Na & I & 5688.205 & -0.450 &2.104 &108.4 & 5.98 &124.5 &6.14 \\ 
 Ni & I & 5587.853 & -2.140 &1.935 & 73.5 & 5.57 & 64.3 &6.20 \\ 
 Ni & I & 5614.768 & -0.508 &4.154 & 34.4 & 5.53 & 45.4 &6.30 \\ 
 Ni & I & 5682.198 & -0.470 &4.105 & 44.8 & 5.67 & 55.3 &6.32 \\ 
 Ni & I & 5748.346 & -3.260 &1.676 & 51.5 & 5.83 & 29.5 &6.25 \\ 
 Ni & I & 5760.828 & -0.800 &4.105 & 36.7 & 5.81 & 36.2 &6.28 \\ 
 Ni & I & 5805.213 & -0.640 &4.167 & 34.6 & 5.66 & 41.8 &6.28 \\ 
 Ni & I & 5831.593 & -1.079 &4.167 & 27.9 & 5.95 & 25.9 &6.41 \\ 
 Ni & I & 5846.986 & -3.210 &1.676 & 36.8 & 5.45 & 22.6 &6.02 \\ 
 Ni & I & 5857.746 & -0.636 &4.167 & 38.0 & 5.75 & 58.9 &6.75 \\ 
 Ni & I & 6007.306 & -3.330 &1.676 & 41.4 & 5.66 & 26.4 &6.23 \\ 
 Ni & I & 6086.276 & -0.530 &4.266 & 43.2 & 5.85 & 43.9 &6.29 \\ 
 Ni & I & 6223.981 & -0.910 &4.105 & 29.8 & 5.73 & 28.1 &6.19 \\ 
 Ni & I & 6314.653 & -1.770 &1.935 & 88.4 & 5.47 & 76.7 &6.00 \\ 
 Ni & I & 6322.164 & -1.170 &4.154 & 17.0 & 5.69 & 18.2 &6.23 \\ 
 Ni & I & 6327.593 & -3.150 &1.676 & 57.6 & 5.81 & 38.6 &6.31 \\ 
 Ni & I & 6360.808 & -1.279 &4.167 & 16.7 & 5.80 &----- &----- \\ 
 Ni & I & 6384.663 & -1.130 &4.154 & 22.9 & 5.82 & 23.9 &6.35 \\ 
 Ni & I & 6414.581 & -1.180 &4.154 & 18.7 & 5.75 & 19.0 &6.26 \\ 
 Ni & I & 6482.796 & -2.630 &1.935 & 64.9 & 5.74 & 41.7 &6.10 \\ 
 Ni & I & 6532.871 & -3.390 &1.935 & 26.7 & 5.66 & 13.7 &6.14 \\ 
 Ni & I & 6586.308 & -2.810 &1.951 & 62.2 & 5.87 & 45.9 &6.38 \\ 
 Ni & I & 6635.118 & -0.820 &4.419 & 21.4 & 5.76 & 26.2 &6.34 \\ 
 Ni & I & 6643.629 & -2.300 &1.676 &114.4 & 6.19 & 96.2 &6.60 \\ 
 Ni & I & 6767.768 & -2.170 &1.826 &106.9 & 6.07 & 81.6 &6.37 \\ 
 Ni & I & 6772.313 & -0.980 &3.658 & 54.5 & 5.82 & 50.3 &6.24 \\ 
 Sc &II & 5641.001 & -1.131 &1.500 & 53.0 & 2.61 & 43.3 &3.39 \\ 
 Sc &II & 5667.149 & -1.309 &1.500 & 56.0 & 2.85 & 32.6 &3.34 \\ 
 Sc &II & 5669.042 & -1.200 &1.500 & 64.8 & 2.94 & 35.4 &3.29 \\ 
 Sc &II & 5684.202 & -1.074 &1.507 & 60.8 & 2.73 &----- &----- \\ 
 Sc &II & 6604.601 & -1.309 &1.357 & 69.5 & 2.91 & 34.9 &3.20 \\ 
 Si & I & 5665.555 & -1.750 &4.920 & 34.8 & 6.98 & 38.0 &7.44 \\ 
 Si & I & 5684.484 & -1.732 &4.954 & 51.3 & 6.95 & 65.7 &7.55 \\ 
 Si & I & 5690.425 & -1.769 &4.930 & 38.5 & 6.89 & 50.2 &7.49 \\ 
 Si & I & 5701.104 & -1.581 &4.930 & 37.3 & 7.05 & 37.2 &7.45 \\ 
 Si & I & 5948.541 & -0.780 &5.082 & 75.0 & 7.10 & 85.9 &7.50 \\ 
 Si & I & 6243.815 & -1.242 &5.616 & 36.3 & 7.10 & 48.3 &7.65 \\ 
 Si & I & 6244.466 & -1.093 &5.616 & 39.3 & 7.19 & 46.1 &7.67 \\ 
 Ti & I & 5662.150 &  0.010 &2.318 & 31.9 & 4.41 & 24.0 &4.86 \\ 
 Ti & I & 5739.978 & -0.670 &2.236 & 16.4 & 4.58 &  6.4 &4.78 \\ 
 Ti & I & 5766.359 &  0.389 &3.294 & 13.9 & 4.62 &  8.4 &4.87 \\ 
 Ti & I & 5866.451 & -0.840 &1.067 & 73.7 & 4.71 & 48.8 &5.02 \\ 
 Ti & I & 5899.294 & -1.154 &1.053 & 58.6 & 4.66 & 36.8 &5.07 \\ 
 Ti & I & 5903.315 & -2.145 &1.067 & 17.0 & 4.73 &----- &----- \\ 
 Ti & I & 5922.109 & -1.466 &1.046 & 28.4 & 4.33 & 19.3 &4.96 \\ 
 Ti & I & 5941.751 & -1.510 &1.053 & 28.9 & 4.39 &----- &----- \\ 
 Ti & I & 5953.160 & -0.329 &1.887 & 53.9 & 4.71 & 35.7 &5.05 \\ 
 Ti & I & 5965.828 & -0.409 &1.879 & 34.3 & 4.36 & 30.0 &5.00 \\ 
 Ti & I & 6064.626 & -1.944 &1.046 & 25.0 & 4.72 &  8.1 &4.98 \\ 
 \hline 
\end{tabular} 
\end{center} 
\end{table*} 
\addtocounter{table}{-1} 
\begin{table*} 
\caption{Adopted line list and atomic parameters. The measured equivalent width and the
corresponding abundances obtained for each line are also reported for both Pal1-I
and the Sun (continued).} 
\begin{center} 
\begin{tabular}{llrrrrrrr} 
\hline 
Element & Ion  & $\lambda$ & log gf  & $\chi$      & EW       &$\epsilon$ & EW       & $\epsilon$\\
        &      & ($\AA$)   &         & eV          & (m$\AA$) &           & (m$\AA$) &           \\
        &      &           &         &             & \object{Pal1-I}         &          & \object{Sun}\\ 
\hline
Ti & I & 6091.171 & -0.423 &2.267 & 18.9 & 4.43 & 14.4 &4.95 \\ 
 Ti & I & 6258.102 & -0.355 &1.443 & 75.2 & 4.64 &----- &----- \\ 
 Ti & I & 6261.098 & -0.479 &1.430 & 84.3 & 4.95 & 50.3 &5.01 \\ 
 Ti & I & 6556.062 & -1.074 &1.460 & 27.4 & 4.35 &----- &----- \\ 
 Ti & I & 6599.105 & -2.085 &0.900 & 28.7 & 4.74 &----- &----- \\ 
 Ti & I & 6743.122 & -1.630 &0.900 & 35.6 & 4.42 & 19.0 &4.92 \\ 
  V & I & 5670.853 & -0.420 &1.081 & 28.5 & 3.32 & 16.5 &3.86 \\ 
  V & I & 5727.048 & -0.012 &1.081 & 55.1 & 3.48 & 38.6 &3.99 \\ 
  V & I & 5727.652 & -0.870 &1.051 & 12.1 & 3.25 &  6.5 &3.81 \\ 
  V & I & 6039.722 & -0.650 &1.064 & 17.8 & 3.23 & 12.3 &3.89 \\ 
  V & I & 6090.214 & -0.062 &1.081 & 48.5 & 3.36 & 33.1 &3.90 \\ 
  V & I & 6216.354 & -1.290 &0.275 & 41.8 & 3.49 & 36.6 &4.39 \\ 
  V & I & 6243.105 & -0.980 &0.301 & 62.1 & 3.63 & 31.3 &3.99 \\ 
  V & I & 6285.150 & -1.510 &0.275 & 36.7 & 3.60 &----- &----- \\ 
  Y &II & 5662.925 &  0.160 &1.944 & 54.6 & 1.86 & 50.5 &2.67 \\ 
 Zn & I & 6362.338 &  0.150 &5.796 & 29.0 & 4.40 & 20.9 &4.51 \\ 
 \hline 
\end{tabular} 
\end{center} 
\end{table*}

\end{document}